# An electrically pumped topological polariton laser.

Philipp Gagel*,[1], Oleg A. Egorov[2], Franciszek Dzimira[1], Johannes Beierlein[1], Monika Emmerling[1], Adriana Wolf[1], Fauzia Jabeen[1], Simon Betzold[1], Ulf Peschel[2], Sven Höfling[1], Christian Schneider[3], and Sebastian Klembt[†,1]

[1] *Julius-Maximilians-Universität Würzburg, Physikalisches Institut and Würzburg-Dresden Cluster of Excellence ct.qmat, Lehrstuhl für Technische Physik, Am Hubland, 97074 Würzburg, Germany*
[2] *Institute of Condensed Matter Theory and Optics, Friedrich-Schiller-Universität Jena, Max-Wien Platz 1, 07743, Jena, Germany*
[3] Institute of Physics, University of Oldenburg, D-26129 Oldenburg, Germany

[*] philipp.gagel@uni-wuerzburg.de    [†] sebastian.klembt@uni-wuerzburg.de

## Abstract

With a seminal work of Raghu and Haldane in 2008, concepts of topology have been successfully introduced in a wide range of optical systems, emulating specific lattice Hamiltonians. Certainly, one of the most promising routes to an application of topological photonics in an actual device are topological lasers, where efficient and highly coherent lasing from a topologically non-trivial mode is achieved. While some attempts have been made to excite such structures electrically, the majority of published fundamental experiments use a form of laser excitation.

In this paper, we use a lattice of vertical resonator polariton micropillars to form an exponentially localized topological Su-Schrieffer-Heeger defect. Upon electrical excitation of the p-i-n doped structure, the system unequivocally shows polariton lasing from the topological defect using a carefully placed gold contact. Despite the presence of doping and electrical contacts, the polariton band structure clearly preserves its topological properties. At high excitation power the Mott density is exceeded leading to highly efficient lasing in the weak coupling regime. This work is an important step towards applied topological lasers using vertical resonator microcavity structures.

# Introduction

The concept of a topological insulators being characterized by the protected transport of electrons in edge channels was first introduced in two-dimensional electron systems for the quantum Hall [1-3] and quantum spin Hall phase [4-6]. This way, abstract concepts of topology have proven to serve as new and surprisingly powerful tool to control and manipulate the flow of electrons. In their seminal work, Haldane and Raghu realized that these topological concepts could be transferred to the realm of light, jump-starting the field of topological photonics [7]. Soon after the first theoretical prediction, topological insulators of light were introduced in microwaves [8], the optical domain [9-11] as well as strongly coupled light-matter systems [12]. While the field has grown and matured immensely [13], the majority of works have expertly implemented linear band structures, raising the question: *Where do we go from here?* [14], with a particular focus on promising applications for future technologies. An exciting answer to this question has certainly been the discovery of topological lasers [15-19]. Early works utilized light confinement within the plane, making the extraction of high-power light increasingly difficult. This issue could be addressed by the implementation of topological vertical cavity surface emitting lasers (VCSELs) [20], where the topological transport takes place in-plane, while coherent laser light is emitted out of plane. Other concepts skip propagative concepts entirely, relying on open-Dirac singularities for scale-invariant lasers [21-23]. An obvious step toward a more application-oriented device is the electrical operation. While some attempts have been made on coupled ring lasers [24] and topological lasers with valley edge modes [25], electrically pumped topological lasers using vertical cavities are absent so far. In this paper, we implement an electrically pumped array of coupled semiconductor microlasers. We unequivocally demonstrate (single mode) lasing from a topological mode, located at a domain boundary. The topological defect is implemented in a Su-Schrieffer-Heeger (SSH) structure, revealing the expected exponential localization of the defect. Excellent optical and material quality allows for the lasing onset in the strong light-matter regime [19,26,27] as well as electrically injected polariton lasing [28,29]. A precise optimization of the p-i-n doping of the semiconductor microcavity as well as a precise placement of the electrical contacts, are a promising and crucial step towards large-scale topological laser arrays.

## Results and Discussion

For the realization of the investigated SSH chain, a p-i-n doped microcavity consisting of GaAs/AlAs mirror pairs for the distributed Bragg reflectors (DBRs) hosting four $In_{0.15}Ga_{0.85}As$ quantum wells in an intrinsic GaAs λ-cavity is chosen (see Methods). The vacuum Rabi splitting in the strong coupling regime yields $2\hbar\Omega = (5.5 \pm 0.2)$ meV with a cavity quality-factor of ~6,000. The same wafer material was previously used to demonstrate the first electrically driven polariton laser and additional information can be found there [28].

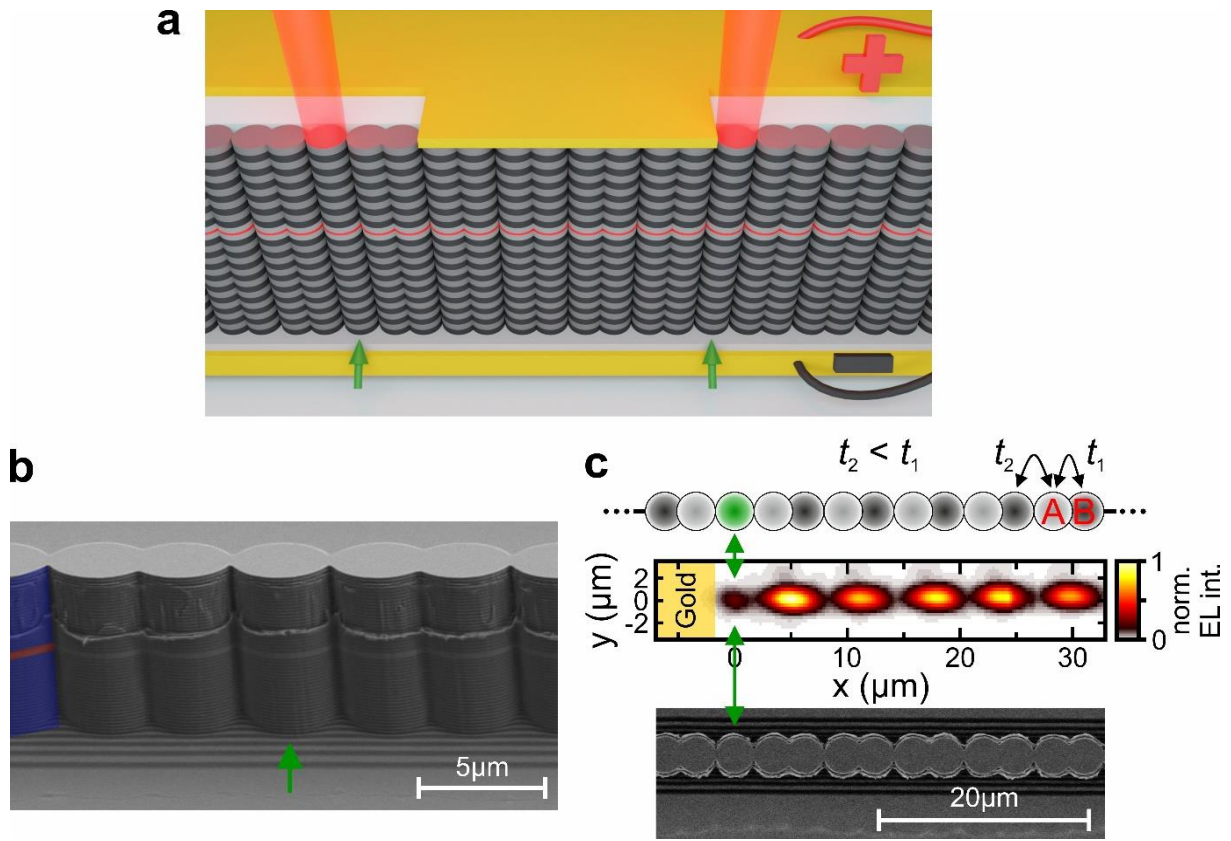

*Figure 1: (a) Sketch of the contacted SSH chain after processing. The domain boundary sites on both sides of the gold contact are marked with a green arrow and lasing of both sites is indicated. (b) Side view SEM image of the SSH chain after the reactive ion etching process with the domain boundary marked by a green arrow. The top and bottom DBRs are highlighted in blue and the cavity region is marked in red. (c) A sketch, a real space EL mode tomography of the binding S-band and a top view SEM image after the etching process of the SSH chain. The focus here is on the right hand side of the gold contact nose in (a) and the position of the domain boundary is marked with a green arrow. The sketch shows the dimers of lattice sites A and B that build up the SSH chain as well as the intra- (inter-) cell coupling constant $t_1$ ($t_2$).*

A sketch of the final device with gold contacts for electrical pumping is depicted in Fig. 1 (a). To achieve this, several process steps are necessary, starting by applying an AuGe-Ni-Au layer sequence as the backside contact via evaporation deposition. Alloying of this contact and also later of the top contact is used to ensure a nearly ohmic contact to the semiconductor material. Next, the SSH chain pattern is defined by means of electron beam lithography and a $BaF_2$-Cr mask is deposited. Subsequently, micropillars with a diameter of $d = 3.5$ µm forming the SSH chain are etched into the microcavity by reactive ion etching. The resulting SSH chain is shown in a side view scanning electron microscope (SEM) image in Fig. 1 (b), where exemplarily the top as well as bottom DBR are highlighted in blue and the cavity layer is marked in red on the left side of the SEM image. Aside from minor sidewall damage to the top DBR, a clean etch profile with a high aspect ratio is obtained, proving the quality of the dry etching method. Following the etching, a planar sample surface is created by spin-coating and baking benzocyclobutene (BCB), a transparent, insulating polymer and the etch mask is removed. Finally, for the top side contact made of Cr-Au another electron beam lithography step is used to define a large area on the BCB for wire bonding individual devices on the sample as well as a small nose, which is the only part actually contacting the SSH chain.

Fig. 1 (c) shows a sketch of an SSH chain consisting of dimer pairs A-B with an intra- (inter-) cell coupling constant $t_1$ ($t_2$). As long as $t_1 \neq t_2$, a bandgap of size $E_{\text{gap}} = 2|t_1 - t_2|$ in the S-band persists. When including a double weakly bound lattice site in the chain (marked in green), a domain boundary defect is created, which hosts a state in the energy bandgap that is topologically protected [19,27,30-35]. In the polaritonic lattice used in this work, the difference in the couplings is realized by varying the overlap of the individual lattice sites, as shown in the top view SEM image in Fig. 1 (c), which clearly reveals the intended dimerized chain geometry. Thereby a greater overlap leads to a greater overlap of the S-orbital wave functions of the individual lattice sites and therefore to a stronger coupling [36,37]. To qualify the overlap of the micropillars, we introduce the overlap parameter $v = a/d$ with *a* denoting the center-to-center distance of two neighboring pillars and *d* denoting the diameter of the pillar. To create a comparatively large bandgap, while ensuring non-vanishing coupling, the overlap parameter is chosen to be $v_2 = 1$ for the inter-cell coupling of the dimers and $v_1 = 0.85$ for the intra-cell coupling [27]. A detailed overview of the implementation of a linear chain SSH Hamiltonian in polaritonic lattices of coupled vertical resonators can be found in the literature [30,32,35]. Fig. 1 (c) also shows an electroluminescence (EL) real-space mode tomography of the SSH chains binding S-band. Here, the right side of the gold contact nose is chosen, where the defect is situated at the very left, directly next to the contact, as sketched in Fig. 1 (a). The tomography shows the typical mode shape of the binding S-band, clearly indicating the position of the coupled dimer pairs.

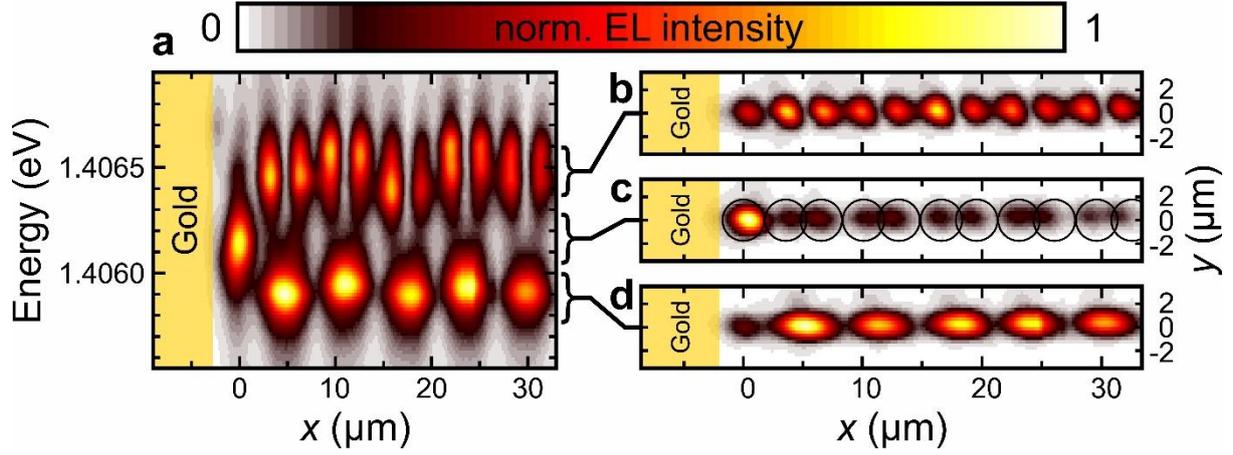

Figure 2: **Real-space EL mode tomography** (a) Energy resolved real space measurement showing the S-band of the SSH chain. A clear topological band gap of $E_{gap} \approx 0.5$ meV persists between the binding and anti-binding S-bands with the emission of the domain boundary site at $x = 0$ lying within the bandgap at $E_{topo} = 1.40618$ eV. (b) - (d) Mode tomography of the (d) binding and (b) anti-binding S-band as well as (c) in the range of the energy band gap. The gap state in (c) clearly dominates the spectrum and is localized at the domain boundary site at $x = 0$. The small contribution of the other lattice sites in (c) is due to the broad line width of the emission of the device below threshold.

One important aspect to consider in the overall design of the device is the positioning of the small contact nose (see Fig. 1 (a)). On the one hand, the contact nose is placed in close proximity to the domain boundary site to ensure efficient electrical pumping. On the other hand, a certain distance to the gold contact can minimize its influence on the band structure and avoid unwanted energy shifts [38]. However, in our experiments we find that the influence on the band structure energy is rather negligible. This is confirmed by the EL real space band structure in Fig. 2 (a), that shows no major energy gradient except for an insignificant variation of the local onsite energy. Therefore, in the following experiments, we concentrate on the right side of the gold contact nose with the domain boundary site located next to the contact as sketched in Fig. 1 (a). However, in the real space spectrum of Fig. 2 (a), the light emission spreads almost homogeneous over tens of micrometers from the metallic contact. Therefore, the topological defects within the SSH chain can in principle be pumped remotely far away from the contact and indeed we find very comparable results for a distance of two micropillars on the left side of the contact nose, which are shown in the Supplementary Information section 1.

With the chain implemented, next we are going to study the band structure of our device and identify the topological gap state therein. Therefore, an energy-resolved real-space tomography of the S-band along the SSH chain is measured. Fig. 2 (a) shows the S-band of the SSH chain with the binding S-band at an energy of $E \approx 1.4059$ eV and the anti-binding S-band centered around $E \approx 1.4065$ eV. Due to the distinct inter- and intra-cell coupling constants

$t_1$ and $t_2$ of the SSH chain, a topological energy band gap of $E_{\text{gap}} \approx 0.5$ meV exists between the two S-bands. The emission of the domain boundary site right next to the gold contact at $x = 0$ µm shows its emission at $E_{topo} \approx 1.4062$ eV, unambiguously falling in the range of the energy band gap. Note that there is a negligible deviation in the onsite emission energy from dimer to dimer in the binding as well as anti-binding S-band, while there is no overall potential gradient. Any emission on the left side of the domain boundary is blocked by the gold contact, whose position is indicated in all panels of Fig. 2. The energy cuts of the real space mode tomography in Fig. 2 (b) – (d) show the (b) anti-binding, (d) binding S-band and (c) the mode located in the energy range of the band gap. The underlying dimer structure of the SSH chain is resolved in the binding S-band in panel (d), whereas in contrast, the anti-binding mode in panel (b) depicts roughly the position of each individual pillar of the chain. The small contribution of the domain boundary site at $x = y = 0$ to both bands is due to its emission line width below the lasing threshold. In panel (c), the position of the underlying SSH chain is indicated to demonstrate that the spectrum within the energy band gap is clearly dominated by the emission from the domain boundary site. Thus, the real-space mode tomography unequivocally proves the implementation of a SSH chain with an energy gap and a topological protected gap state at $E_{topo} \approx 1.4062$ eV, localized at the domain boundary site.

Next, we increase the direct current (DC) to perform a power series and present the resulting input-output characteristics of the gap state in Fig. 3, where remarkably a two-threshold behavior is found. Therefore, panels (a), (b), and (c) show an energy resolved real-space measurement at a current (a) well below the first threshold ($I = 0.02$ mA), (b) between the two thresholds ($I = 5.28$ mA), and (c) above the second threshold ($I = 9.56$ mA). From (a) to (b), a small blue shift for the gap state can be observed (see also Fig. 3 (e)) and the output intensity concentrates predominantly at the domain boundary at $x = 0$. By further increasing the excitation current, in (c) a clear red shift of the modes exists in the spectrum (see also Fig. 3 (e)) and the line width drops significantly, leading to a more pronounced band gap between the binding and anti-binding S-band. A more detailed analysis of the gap state emission by fitting a Lorentz function to the peak of each individual measurement produces the input-output characteristic shown in Fig. 3 (d). To better illustrate the nonlinearities associated with the two-threshold behavior, fits to the linear sections of the graph are depicted in blue. The intersections of the fits at $I_{th}^{Pol} = 1.15$ mA and $I_{th}^{Photo} = 6.65$ mA are marked by gray areas and go together with a first moderate and a second more pronounced decrease in line width by a factor of four, respectively. The characteristic laser behavior at $I_{th}^{Pol}$ is accompanied by a blue shift of the energy by $\Delta E \approx 0.2$ meV, as shown in Fig. 3 (e). We attribute the appearance of these three typical features to the onset of polariton lasing, where the coherent emission naturally affects the polariton relaxation, which is stimulated and thus more efficient than the spontaneous processes below the threshold. At the same time, the blue shift results from the increasing polariton-polariton interaction with increasing occupation density. After reaching about four times $I_{th}^{Pol}$, the Mott density in the quantum wells is exceeded, leading to a breakdown of the strong coupling regime [39]. Simultaneously, a pronounced red shift of the energy occurs in Fig. 3 (e), which is due to the increased heating of the device caused by the high current. We attribute this second threshold to the transition to the photon lasing regime. Here, the nonlinearity shows a dramatic increase of the output power by more than three orders of magnitude, which is further discussed in the Supplementary Information section 2. Such a two-threshold behavior as observed in Fig. 3 is not unusual for polaritons and has been widely discussed in the literature [26,40-43].

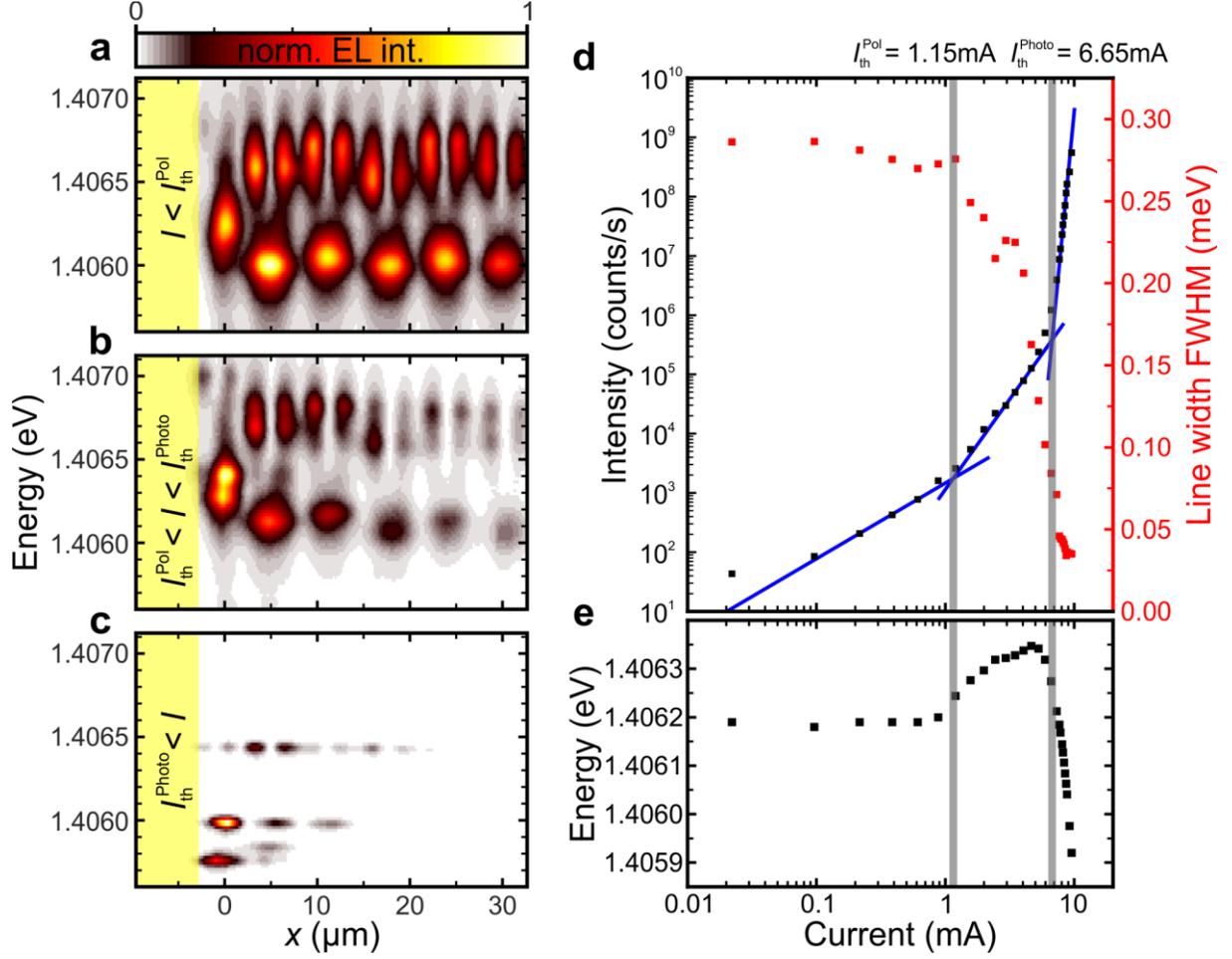

Figure 3: **Input-output characteristic of the topological gap state** (a) - (c) Energy resolved real space measurement of the S-band of the SSH chain at a power (a) below the first threshold, (b) between the two thresholds and (c) above the second threshold. (d) Input-output characteristic of the gap state showing a two-threshold behavior. We assign the first one to the onset of polariton lasing due to the blue shift found in (e) and the second one to the transition to the photon lasing regime due to the increased heating of the device under the high current that causes the red shift in (e). The blue lines are fits to the linear sections of the graph and help to indicate the two nonlinearities. (e) Energy shift of the gap state with increasing current.

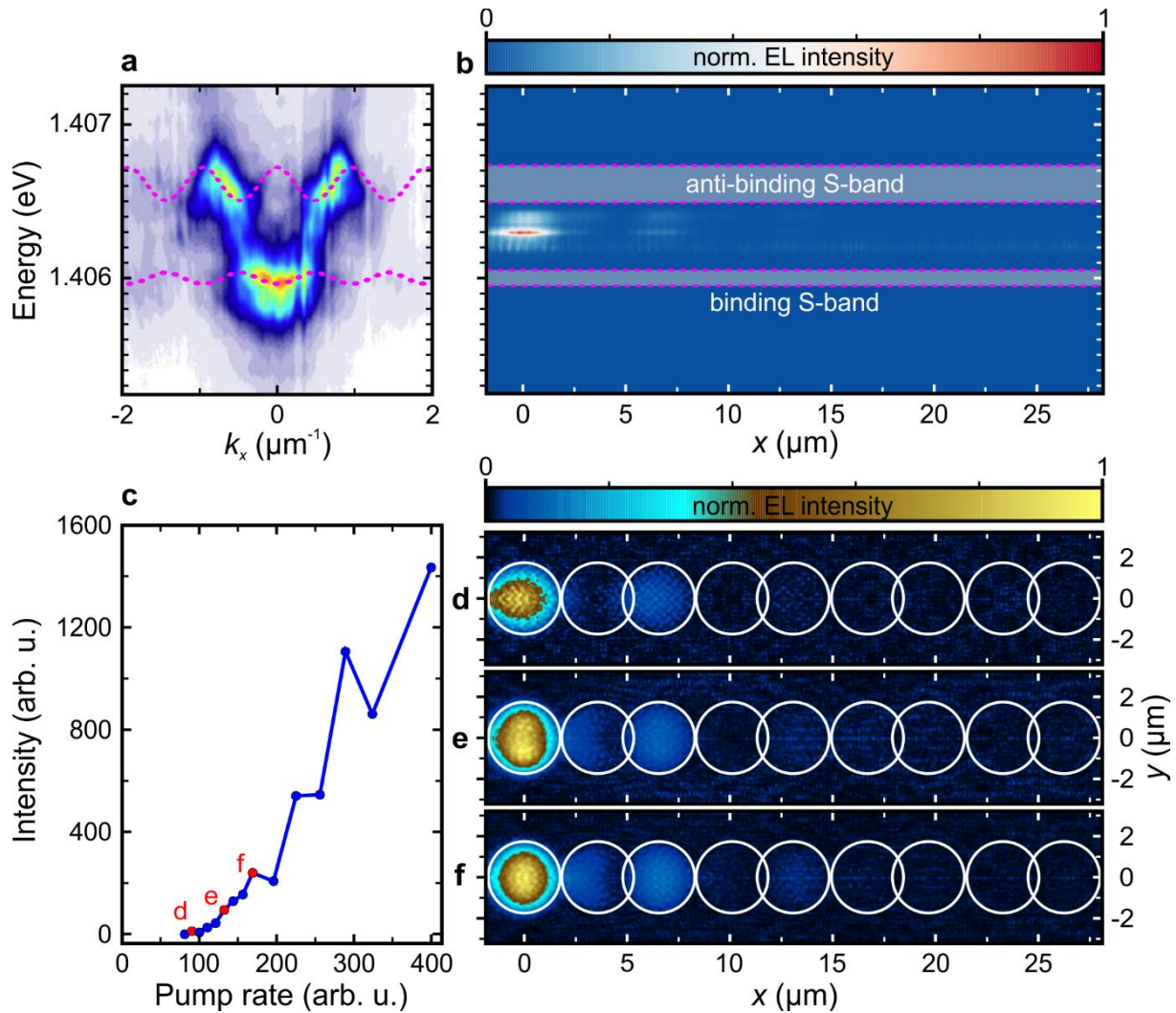

*Figure 4: **Numerical simulation of polariton lasing from the defect.** (a) Dispersion relation of the SSH chain calculated in the framework of the exciton-polariton basis model with an effective potential (dashed lines) in comparison with the experimentally measured spectra of the device. (b) Topological defect state calculated in the modified Gross-Pitaevskii approach above the condensation threshold for the pump rate equal to 169 arb.u.. The energy of the defect state is within the topological gap between the two S-bands of the SSH chain. (c) Input-output characteristic of the emission from the topological defect showing a distinct non-linear behavior with the threshold at a pumping rate of 81 arb.u. The electrical pumping was simulated by injecting of the particle into an incoherent reservoir in the pillars under the contact. (d-f) Real space representation of the defect mode (normalized intensity) calculated above the condensation threshold for the pumping rates 90.25, 132.25 and 169 arb.u., respectively.*

To shed light on the lasing behavior, we perform numerical simulations using a modified Gross-Pitaevskii approach [27,44] derived for the polaritons in the vicinity of the lower-polariton branch and coupled to an incoherent reservoir of excitons. The model assumes realistic experimental input parameters such as the geometry of the trapping potential, the Rabi splitting, the exciton-photon detuning, the photon effective mass and the photon lifetime. More information about the theoretical model can be found in the Supplementary Information section 3. First, we calculate the dispersion relation of the respective 2D Bloch modes of the SSH chain within the photon-exciton basis. We solve a standard eigenvalue problem with an effective 2D photonic potential [12], implemented by the etched pillars. Fig. 4 (a) shows both the experimental and the calculated energy dispersion (dashed lines) of the binding and anti-binding S-bands separated by the energy bandgap of the SSH chain. We then proceed with the simulation of the nonlinear exciton-polariton dynamics above the lasing threshold within the modified Gross-Pitaevskii approach. Assuming direct pumping of the incoherent exciton reservoir by electric current, the pump spot is aligned with the gold contact adjacent to the domain boundary site. With increasing pumping rate, clear polariton lasing from the defect mode within the topological energy gap arises in Fig. 4 (b). Fig. 4 (c) shows the nonlinear increase of the output intensity above the lasing threshold of the defect mode in a lin-lin plot, which refers to the experimental lasing threshold $I_{\text{th}}^{\text{Pol}} = 1.15$ mA. In addition, the exponential decay of the defect mode into the bulk is well reproduced, as exemplarily shown by the spatial profiles in Fig. 4 (d) - (f). It is worth mentioning that this intensity pattern, typical for the topological defect mode, is very robust to variations of system parameters and exists from the condensation threshold up to very high pumping rates. Overall, the calculations are in good agreement with the experimental data of the polariton lasing presented in Fig. 3.

To further investigate the topological nature of the gap state hosted by the domain boundary site, a real space mode tomography is performed just above the second threshold $I_{\text{th}}^{\text{Phot}}$ at a current of $I = 6.7$ mA. The resulting mode distribution within the energy gap of the SSH chain is shown in Fig. 5 (b). To illustrate the position of the individual pillars in Fig. 5 (a), a sketch of the SSH chain is placed directly above the mode tomography. Here, the two sub lattices A and B of the SSH chain are indicated as well as the domain boundary site, which is highlighted in green and belongs to the sub lattice B. The main emission within the energy bandgap originates as expected from the domain boundary site at $x = y = 0$ as shown in Fig. 5 (b) and can be attributed to the topologically protected gap state. Next, the emission in (b) is integrated along the *y*-direction and the normalized intensity distribution is plotted against *x* in Fig. 5 (c). The individual maxima in the graph all belong to the sub-lattice B and form an exponential decay that reaches the characteristic value of $e^{-1}$ already after $x = 2.75$ μm. This indicates the strong localization of the topological gap state at the domain boundary site. The restriction to the sub-lattice of the domain boundary site and the exponential localization are characteristic features for a topological gap state in a SSH chain and highlight the topological nature of the gap state laser emission [31,32,35,45].

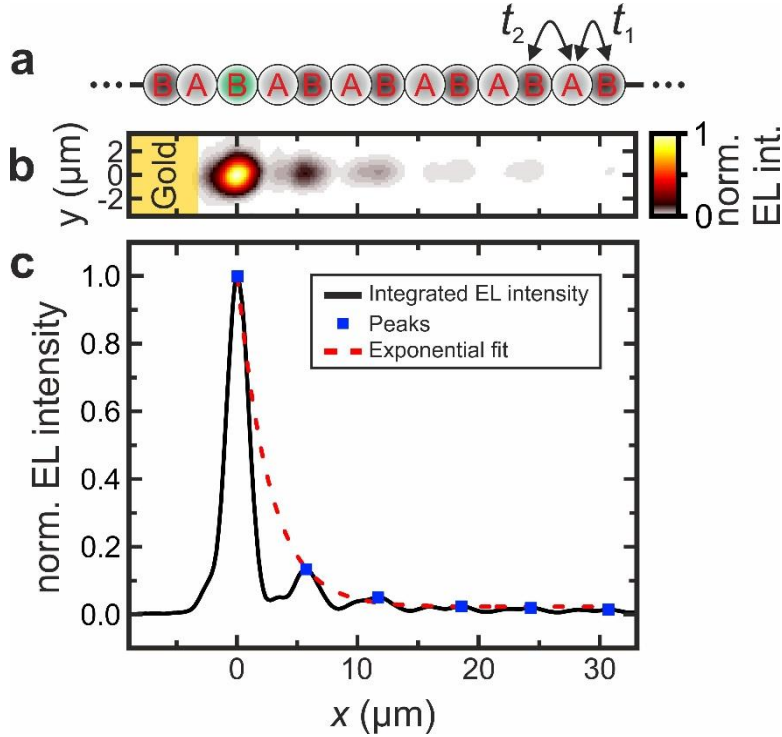

*Figure 5: **Spatial exponential decay of the output intensity of the gap state above threshold** (a) Sketch of the SSH chain with coupling constants $t_1$ and $t_2$. The two sub lattices A and B are indicated and the domain boundary site, that is attributed to the sub-lattice B, is highlighted in green. (b) Mode tomography in the energy range of the band gap above threshold. The gap state is localized at the domain boundary site at $x = y = 0$. (c) Normalized output intensity integrated along the y-direction in (b). The individual peaks found in the spectrum are all located on the sub-lattice B, forming an exponential decay.*

## Conclusion

In summary, we have exploited all electrical DC pumping to demonstrate polariton lasing of a topologically protected gap state in a one-dimensional Su-Schrieffer-Heeger chain made out of coupled polariton microresonators. Using tomographic spectroscopy techniques, we unambiguously prove the existence of the topological gap state in our system under DC pumping. A carefully placed gold contact in close proximity to the defect site allows the defect site predominantly to be pumped and lasing to be achieved without major distortion of the band structure. A power series shows a two-threshold behavior, which is interpreted as the onset of polariton lasing and followed by the breakdown of the strong coupling regime exceeding the Mott density in the quantum wells as the transition to the photon lasing regime. The topological nature of the defect persists above the thresholds, which is emphasized by the exponential localization of the defect emission intensity at the domain boundary site in a real-space mode tomography. Numerical simulations fully reproduce the polariton lasing onset and the modal structure of the topological laser mode. Our findings are an important step towards the development of robust electrically pumped topological polariton and photon lasing devices. In particular, vertical cavity laser approaches provide good light extraction, modulation speed and a compact design, as well as generally good characteristics for applications. By extending our system to two dimensions, even scaling of the output power by coupling via topologically protected states as well as room temperature operation is possible [20,24].

## Methods

**Sample.** The microcavity sample is grown on an n-doped (100) GaAs substrate by molecular beam epitaxy and consists of 27 (23) AlAs/GaAs mirror pairs for the bottom (top) DBR. The Si n-type (C p-type) doping of the bottom (top) DBR is reduced stepwise from $3 \cdot 10^{18}$ cm$^{-3}$ to $1 \cdot 10^{18}$ cm$^{-3}$ towards the intrinsic cavity. To ensure a good semiconductor-metal contact, the two top most mirror pairs of the upper DBR are heavily p-doped with a concentration of $2 \cdot 10^{19}$ cm$^{-3}$. Delta doping layers with a sheet density of $10^{12}$ cm$^{-2}$ are implemented at every second mirror pair interface to reduce the resistivity of the DBRs. The GaAs λ-cavity contains $4 \times 8$ nm wide $In_{0.15}Ga_{0.85}As$ quantum wells with a 6nm GaAs barrier placed at the center of the cavity, where the electric field distribution exhibits an anti-node. In this setting, the strong coupling regime is reached with a vacuum Rabi splitting of $2\hbar\Omega = (5.5 \pm 0.2)$ meV measured by reflection measurements on the unetched sample. A quality factor of $6320 \pm 60$ is determined from photo-luminescence spectra on a far red detuned $d = 2$ μm pillar. The investigated SSH chain sits at an exciton-photon detuning of approximately $-19.8$ meV.

**Measurement:** The sample is held at 4 K in a Janis ST-500 continuous-flow helium cryostat with electrical feedthrough and the current flowing through the sample is measured by the voltage drop on a 1 kΩ series resistor. A Mitutoyo objective with a numerical aperture of 0.42 and a magnification of 20x is focused on the sample surface to collect the electro-luminescence emission of the device. The emitted photons are passed through a spectroscopy setup with a movable projection lens enabling to scan the real space and therefore perform mode tomographies. To analyze the spectrum of the light, a high-resolution Czerny-Turner spectrometer (Andor Shamrock SR-750) with a minimum resolution of about 20 µeV is used which focuses the light on a Peltier cooled ($T = -70$ °C) CCD camera (Andor ikon-M).

## Acknowledgements

We gratefully acknowledge the financial support by the state of Bavaria, the German Research Foundation (DFG), within the projects Schn1376/13.1, and KL3124/3.1. The Würzburg Group acknowledges financial support by the DFG under Germany's Excellence Strategy - EXC2147 *ct.qmat* (Project No. 390 858 490) and is grateful for support by the state of Bavaria.

## Competing Interests

The authors declare no competing interests.

## Data availability

The data that support the findings of this study are available from the corresponding author upon reasonable request.

# Supplementary Information for

# An electrically pumped topological polariton laser.

Philipp Gagel*,1, Oleg A. Egorov2, Franciszek Dzimira1, Johannes Beierlein1, Monika Emmerling1, Adriana Wolf1, Fauzia Jabeen1, Simon Betzold1, Ulf Peschel2, Sven Höfling1, Christian Schneider3, and Sebastian Klembt†,1

1 *Julius-Maximilians-Universität Würzburg, Physikalisches Institut and Würzburg-Dresden Cluster of Excellence ct.qmat, Lehrstuhl für Technische Physik, Am Hubland, 97074 Würzburg, Germany*
2 *Institute of Condensed Matter Theory and Optics, Friedrich-Schiller-Universität Jena, Max-Wien Platz 1, 07743, Jena, Germany*
3 Institute of Physics, University of Oldenburg, D-26129 Oldenburg, Germany

* philipp.gagel@uni-wuerzburg.de   † sebastian.klembt@uni-wuerzburg.de

**Section 1: Topological mode and lasing on the left side of the gold contact**

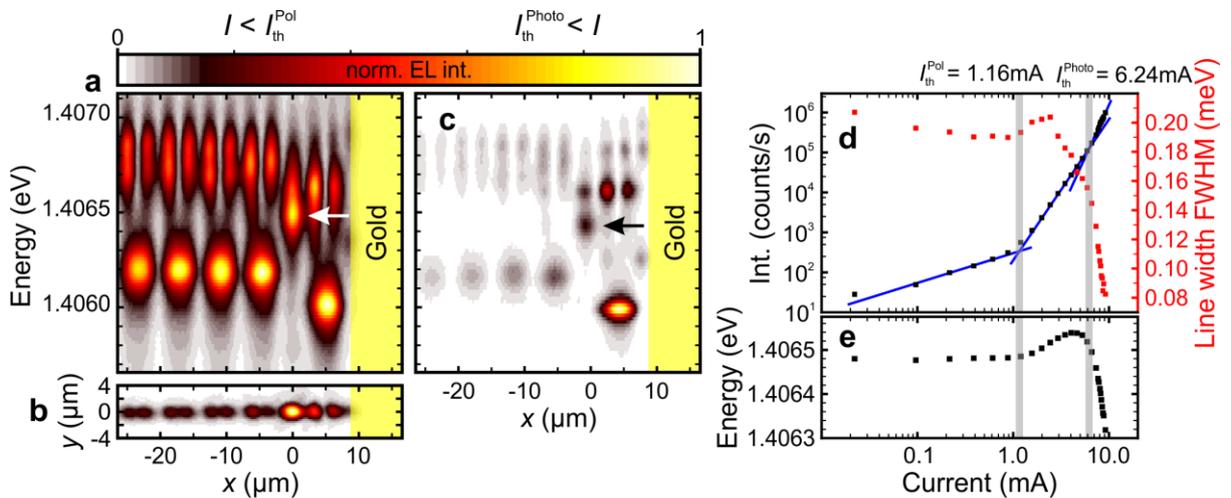

*Figure S1: (a) and (c) Real space spectrum of the SSH chains S-band on the left hand side of the gold contact (a) below $I_{th}^{Pol}$ and (c) above the second threshold $I_{th}^{Photo}$. An arrow marks the topological gap mode located at the position of the domain boundary site (x=0). (b) Real space mode tomography in the energy range of the band gap. The intensity distribution in the energy gap is dominated by the emission of the domain boundary site. (d) Input-output characteristic of the topological gap mode showing a two-threshold behavior. (e) Energy shift of the topological gap mode depending on the input current.*

On the left side of the gold contact, where two lattice sites separate the domain boundary site from the gold contact, we find results similar to those discussed in the main text and that are summarized in Fig. S1. The real space spectrum at a current far below the first threshold $I_{th}^{Pol}$ in Fig. S1 (a) shows both the binding and anti-binding S-bands of the SSH chain on the left side of the gold contact. Here, at the position of the domain boundary site at $x = 0$ and marked by a white arrow, a topological mode exists within the band gap. The real space mode tomography in the energy range of the band gap in Fig. S1 (b) clearly shows that the intensity distribution is dominated by the emission of the domain boundary site, further proofing the topological nature of this mode. Fig. S1 (c) presents a real space spectrum at a current above the second threshold $I_{th}^{Photo}$. In this case, the highest intensity is found in the binding S-band, but still at the sites closest to the gold contact in agreement with Fig. 3 of the main text. Analyzing the behavior of the gap mode (black arrow) by fitting Lorentz curves to the extracted line profiles of each individual measurement, the input-output characteristic is extracted as shown in Fig. S1 (d) and (e). Here, similar to the right side of the gold contact, we find a two-threshold behavior with nonlinearities at $I_{th}^{Pol} = 1.16$ mA and $I_{th}^{Photo} = 6.24$ mA. The first threshold shows a moderate drop in line width after a slight increase and a continuous blue shift (Fig. S1 (e)), clear attributes for the onset of polariton lasing from the topological gap mode. After reaching the Mott density at the second threshold and therefore creating photon lasing in the weak coupling regime, a clear drop in line width of a factor of two occurs. Due to the high current flowing through the sample, it heats up, causing the red shift after the second threshold as shown in Fig. S1 (e).

**Section 2: Nonlinearity in the output intensity during the transition to the weak coupling regime**

The two-threshold behavior with its distinctly different slopes in Fig. 3 of the main text is only clearly visible, when using a logarithmic plot. For comparison, a linear plot of the experimental EL characteristic as a function of the input is shown in Fig. S2. Here, the photon lasing clearly dominates visually and the transition is facilitated by an interplay of carries dephasing as well as the quantum confined Stark effect [S1] and sample heating (see Fig. 3 (e)), causing the exciton to redshift. After crossing the Mott density, the gain overlap between photon and bandgap emission leads to the dominant photon lasing, with the intensity increasing rapidly by four orders of magnitude (see Fig. 3 (d)).

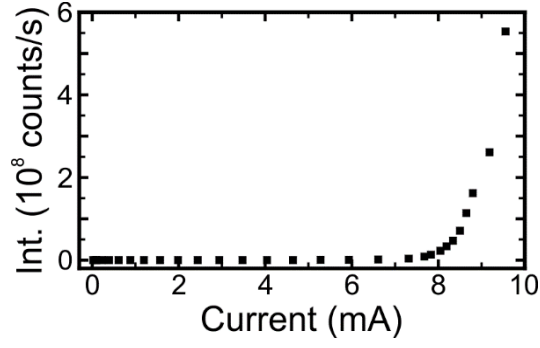

*Figure S2: EL input-output characteristic of the gap state. By using a lin-lin plot only the nonlinearity corresponding to the transition to the weak coupling regime is visible. The data is the same as in Fig. 3 of the main text.*

**Section 3: Gross-Pitaevskii model above the condensation threshold**

For the theoretical investigation of the exciton-polariton behavior in the electrically pumped SSH chain, we use a generalized approach based on the Gross-Pitaevskii (GP) model [S2], accounting a collective wave function of lower-branch polaritons $\Psi(r,t)$ and a high-energy exciton reservoir density $n_R(r,t)$

$$i\hbar \frac{\partial \Psi(r,t)}{\partial t} = \left[ -\frac{\hbar^2}{2m_{\text{eff}}} \nabla^2 - \frac{i\hbar\gamma_C}{2} + V_{\text{ext}}(r) + g_C(r)|\Psi(r,t)|^2 \right.$$

$$\left. + \left( g_R(r) + \frac{i\hbar R(r)}{2} \right) n_R(r,t) \right] \Psi(r,t) + i\hbar \frac{d\Psi_{\text{st}}(r,t)}{dt}$$

$$\frac{\partial n_R(r,t)}{\partial t} = -(\gamma_R + R(r)|\Psi(r,t)|^2) n_R(r,t) + P(r)$$

where $r \equiv \{x,y\}$ describes the position in the plane of the microcavity. The model explicitly accounts for fluctuations derived within the truncated Wigner approximation [S3,S4], giving $d\Psi_{\text{st}}(r_l) = \sqrt{(\gamma_C + R(r_l)n_R(r_l))/(4\delta x \delta y)}\, dW_l$. Here, $dW_l$ is a Gaussian random variable characterized by the correlation functions $\langle dW_l^* dW_j \rangle = 2\delta_{l,j} dt$ and $\langle dW_l dW_j \rangle = 0$ where $l$, $j$ are discretization indices for the spatial coordinate $r \equiv \{x,y\}$.

The following system parameters were used for the calculations: the Rabi splitting $2\hbar\Omega = 5.5$ meV, photon-exciton detuning $\hbar(\omega_{\text{cavity}} - \omega_{\text{exciton}}) \equiv \hbar\Delta_0 = -19.8$ meV, and the effective mass of intracavity photons $m_C = 30.4 \cdot 10^{-6} m_e$ with the free electron mass $m_e$. Then, the effective mass of lower-branch polaritons can be estimated by $m_{eff} = 2m_C\sqrt{\Delta_0^2 + 4\Omega^2}/(\sqrt{\Delta_0^2 + 4\Omega^2} - \Delta_0)$. The effective trapping potential for polaritons is given by the expression $V_{ext}(r) = \frac{1}{2}\left(V(r) + \sqrt{(\Delta_0 - V(r))^2 + 4\Omega^2} - \sqrt{\Delta_0^2 + 4\Omega^2}\right)$, where the external potential for the intracavity photons is determined by the SSH chain consisting of the mesas in the form of super Gauss $V(r) = \sum_i V_0 \exp\left(-\frac{|r-r_i|^6}{d^6}\right)$ with the potential depth $V_0 = 18$ meV and pillars diameters $d = 3.6$ μm (centered at $r_i$). The longer- and shorter-separations between the center of pillars in the SSH chain are 2.975 μm and 3.57 μm, respectively, whereas the period of chain is 6.545 μm.

Since the depth of the external potential is comparable with Rabi splitting, the content of the photonic and excitonic components in polaritons cannot be longer considered as spatially homogeneous. To take into account the spatial variation, we scale the key system parameters, such as stimulated scattering $R(r)$ and polariton-polariton interaction $g_C(r)$ and $g_R(r)$, in accordance with the local fraction of the excitonic component by rescaling them with the Hopfield coefficients:

$$|C(r)|^2 = \frac{1}{2}\left(1 - \frac{\Delta_0 - V(r)}{\sqrt{(\Delta_0 - V(r))^2 + 4\Omega^2}}\right), \quad |X(r)|^2 = 1 - |C(r)|^2.$$

Then, the stimulated scattering from the reservoir to (coherent) polaritons becomes $R(r) = R_0|X(r)|^2$ with the fitting parameter $\hbar R_0 = 0.01$ meV μm². The strengths of polariton-polariton and polariton-reservoir interactions are characterized by the expressions $g_C(r) = g|X(r)|^4$ and $g_R(r) = g|X(r)|^2$, respectively, where $\hbar g = 0.05$ meV μm² describes the effective interaction between bare excitons. The constants $\hbar\gamma_C = 0.074$ meV and $\hbar\gamma_R =$

0.2 meV stand for the decay rates of condensed polaritons and reservoir, respectively, outside the metallic contact. In the model we assume that the polaritons trapped under metallic contacts have slightly smaller losses $\hbar\gamma_C = 0.034$ meV due to back reflection of the photons from the metallic contacts. These losses are compensated by an external off-resonant time-independent electrical pump with the injection rate $P(\boldsymbol{r})$.